# Space, matter and topology *

**Manuel Asorey**

Departamento de Física Teórica. Facultad de Ciencias.
Universidad de Zaragoza.
50009 Zaragoza, Spain

*asorey@unizar.es*

**An old branch of mathematics, Topology, has opened the road to the discovery of new phases of matter. A hidden topology in the energy spectrum is the key for novel conducting/insulating properties of topological matter.**

In the early 1960s George Gamow wrote a popular book [1] that contained a very suggestive observation about the effectiveness of pure mathematics in physics. As a conclusion he remarked that "only number theory and topology (analysis situs) still remain purely mathematical disciplines without any application to physics. Could it be that they will be called to help in our further understanding of the riddles of Nature?" Fifty years later, the applications of topology in physics are so numerous that they cannot be covered in a short note. Gamow surely never dreamt that newly discovered phases of matter would be labelled with topological names. Topological phases, topological insulators, topological superconductors and topological semimetals — these are just a few examples of what is generally referred to as topological matter.

Although for many years topology has been perhaps under-appreciated as a mathematical tool for physics, it is now becoming relevant for addressing a multitude of issues, such as the stability of subtle physical effects under small perturbations, or the robustness of a physical device that might be used, for instance, in a quantum computer. In this sense, topology is a new mathematical tool that has major relevance for phenomena that require stable, clear and sometimes dichotomic answers.

The history of topology in quantum physics begins in gauge theories, with the emergence of new phenomena like the Aharonov–Bohm effect, magnetic monopoles and gauge anomalies. It has even been conjectured that topology could play a fundamental role in the quark confinement mechanism of Quantum Chromodynamics [2]. But a topological revolution has recently taken place in quantum matter, with many new phases discovered in which topology plays a leading role. A special feature of such topological phases of matter is that they are characterized not by a local order parameter of a spontaneous symmetry-breaking mechanism, but instead by topological invariants that have a global dependence on characteristic parameters of the system.

---

**A hidden topology.**

Topology is the branch of mathematics that deals with properties of spaces that are invariant under smooth continuous deformations. The history of topology starts with Leonard Euler and his topological solution to the Könisberg bridge puzzle, in which he proved that crossing all the seven bridges of the town requires you to cross at least one bridge twice.

Quantum effects due to the topology of physical space can arise only at very low-energy scales. Higher-energy modes are insensitive to space topology and to spatial boundary conditions that manifest only at large spatial (low-energy) scales. For the same reason, most of the information on the topological structure of our universe is encoded in lowest multipoles of the angular spectrum of cosmic microwave background anisotropies [3–6]. However, the emergent topological phases of matter are generated not by the topology of the different material samples, but rather by the topology of the bands of its energy spectrum.

One of its main goals of topology is to classify the different families of spaces that are not equivalent under continuous deformations. If the space is low-dimensional, compact and without boundaries, there are very few non-equivalent types of topological spaces. In one dimension the only possible topology is that of the circle, which is denoted as $S^1$. In two dimensions there is an infinite series of non-equivalent topological spaces: the sphere $S^2$, the sphere with one hand (which is topologically equivalent to the torus $T^2 = S^1 \times S^1$), the sphere with two hands, and so on.

In higher dimensions the simplest topology is that of an $n$-dimensional sphere $S^n$, where $n$ is any natural number. But there are other higher-dimensional spaces with nontrivial topology, such as the Grassmannian $Gr(m, n)$, which is the space of $m$-dimensional subspaces of an $n$-dimensional complex vector space $\mathbb{C}^n$ ($m \leq n$). The Grassmannian is relevant to the classification of topological phases of matter because $Gr(m,n)$ is associated with an $n$-dimensional quantum system with $m$ occupied energy levels.

The classification of all topological spaces with dimensions larger than two is a difficult problem, and a good strategy for tackling such topologies is to define enough topological invariants to discriminate between the different topologies. Another goal of topology is the classification of continuous maps from one space $X$ into another space $Y$ that are not equivalent under continuous deformations. The set of all non-equivalent (homotopy) classes is usually denoted by $[X,Y]$. The homotopy class of a given map in $[X,Y]$ is by definition a topological invariant that cannot be modified by any continuous or smooth perturbation.

**The Pandora's Box of topological matter.**

In crystalline solids and systems with periodic symmetries, the bulk energy spectrum presents a band structure that can be understood in terms of Bloch waves with momenta $k$ confined to a bounded Brillouin zone $T$. Usually this space has a nontrivial topology even if the original physical space is topologically trivial. The interesting connection of quantum matter with topology arises from the rich topological structure of the space of physical states when considered as a Bloch bundle $E(T)$ over the Brillouin zone $T$, where the fibres are the spaces of states with the same Bloch momentum $k$. The $k$-dependent Hamiltonians $H(k)$ acting on the different fibres of the Hermitian Bloch vector bundle $E(T)$ generate the bands of the energy spectrum. If two different Hamiltonians in $E(T)$ can be deformed into each other by continuous perturbations they share some common physical properties that are homotopy invariant. From that perspective what matters is the topological structure of the bundle $E(T)$ and not the particular Hamiltonian of the solid, because any two Hamiltonians acting on the same bundle of states are topologically equivalent.

The robustness provided by topological invariance and some extra discrete symmetries is what motivated interest in the search and discovery of new topological phases of matter. The presence of symmetry-protected topological invariance guarantees the stability of insulating–conducting behaviour under small amounts of disorder or continuous deformations of the system Hamiltonian.

In the integer quantum Hall effect, the Bloch bundle $E(T^2)$ of each electronic band is an $m$-dimensional Hermitian vector bundle over the two-dimensional (2D) Brillouin torus $T^2$, where $m$ is the number of units of external magnetic flux per cell. The classes of Bloch bundles $E(T^2)$ are classified by the homotopy classes of the maps from $T^2$ to the Grassmannian spaces $Gr(m,m+n)$ (with $n > 1$) [7], which are characterized by an integer known as the first Chern class of the bundle. The Chern class of the Bloch bundle $E(T^2)$ of $v$ filled bands is just $v$, and it is the integer character of this topological invariant index that is responsible for the quantization of the Hall conductivity $\sigma = ve^2/h$ in planar Hall samples, in absence of strong electronic self-interactions [8].

To understand the topological complexity of Bloch bundles, let us replace the toric topology of the Brillouin zone with a spherical one. The simplest nontrivial vector bundle over the $S^2$ sphere is the Hopf bundle, which is defined by the pure quantum states of a two-level system. The fibres of the Hopf bundle are made of vectors of the complex space $\mathbb{C}^2$ that represent the same pure quantum state in $Gr(1,2) = S^2$. If we consider only normalized vectors in $\mathbb{C}^2$, the bundle space becomes the $S^3$ sphere and the fibres reduce to $S^1$ circles (Fig. 1). The Hopf bundle explicitly illustrates how a highly nontrivial topological structure is hidden in the space of quantum states, which is in contrast with the trivial structure of $\mathbb{C}^2$ as a Hilbert space. The Hopf bundle has a unit first Chern class of $v = 1$ and can be considered as the atomic bundle that generates all other 2D vector bundles over the sphere $S^2$.

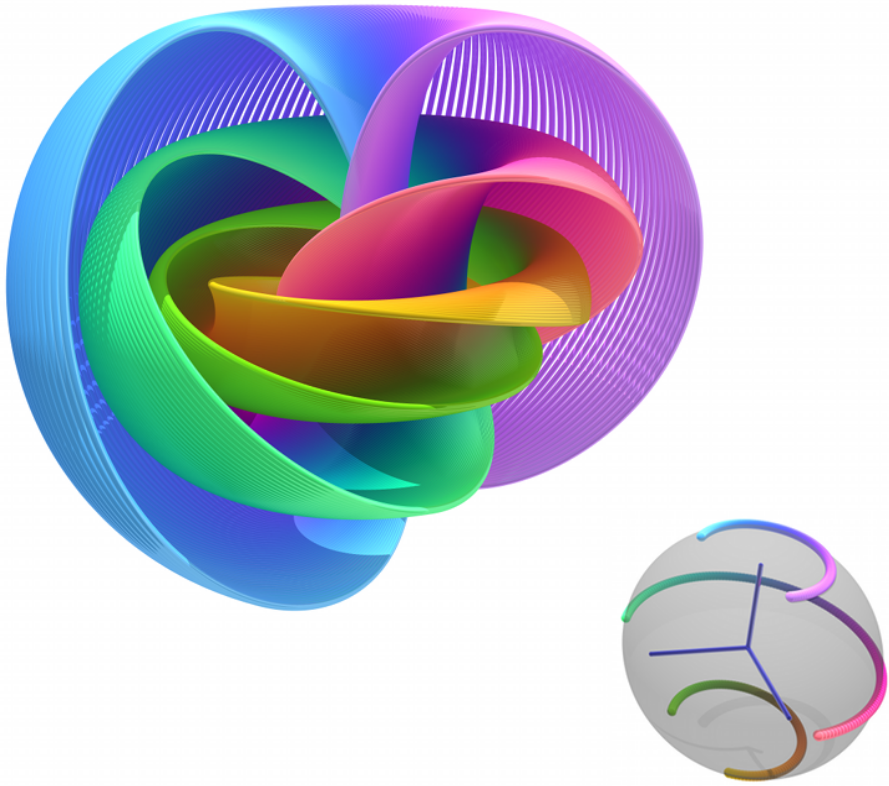

**Figure 1.** Hopf bundle. Stereographic projection of the fibre structure of an $S^3$ sphere and the corresponding Hopf bundle over an $S^2$ sphere. The colour shows to which point on the $S^2$ sphere the distinct circle on the $S^3$ sphere corresponds.

In the case of topological insulators [9–11] with only spin-orbit interactions, the Bloch Hamiltonians $H(k)$ preserve time-reversal symmetry, which imposes additional constraints on the physical states of the Bloch bundle $E(T)$. The homotopy classes of Bloch Hamiltonians that preserve time-reversal invariance in $E(T)$ are different from those of the integer Hall effect [12]. For 2D topological insulators, the classes of time-reversal equivariant maps of the Brillouin zone $T^2$ into the Grassmannian $Gr(m,m+n)$ with $m, n > 1$ are parameterized by a single bit $v$ in the binary group $\mathbb{Z}_2 = \{0,1\}$, which is the topological invariant of the Kane–Mele index $(-1)^v$ (ref. [13]). The $v$-index discriminates between a topological insulator ($v = 1$) and a normal insulator ($v = 0$). In 3D topological insulators the relevant topological invariants are the homotopy classes of time-reversal equivariant maps from the Brillouin zone $T^3$ into the Grassmannian $Gr(m, m+n)$, with $m, n > 1$. They are parameterized by four binary invariants ($v_0, v_1, v_2, v_3$) in $(\mathbb{Z}_2)^4$ (ref. [14]).

Computing the topological indices for a particular Bloch bundle $E(T)$ is generally difficult. However, in some cases it can be obtained by integrating out gauge-invariant local functions that involve the curvature of gauge fields defined on $E(T)$. In the integer Hall effect one can use the gauge field defined by the Berry phase [15] of the occupied states, which is the Fourier–Mukai transform [16] of the external magnetic field. The integer number $v$ of the Hall effect corresponds to the effective magnetic monopole charge of the Berry gauge field enclosed by the torus. For 2D topological insulators this Chern class vanishes, and unfortunately the gauge characterization of the Kane–Mele index $\nu$ is not so simple. However, in three dimensions the index $\nu_0$ can be computed by integrating out the Chern–Simons term of the Berry phase gauge fields [17].

In contrast with the above characterization of topological insulators in terms of the topological properties of their bulk spectrum, there is an alternative characterization based on only the topology of the spectrum of edge states due to the existence of a bulk–boundary correspondence. In finite-size samples the most remarkable change in the energy spectrum is the appearance of gapless edge states localized near the boundaries of the topological insulator closing the gap between two gapped bulk bands. The graphs of the gapless energy levels on the boundary Brillouin zone $S^1$ cross at time-reversal-invariant points due to Kramer's theorem. In the case of 2D topological insulators, the Kane–Mele index $\nu$ corresponds to the sum (modulo 2) of the boundary energy levels per edge that intersect the Fermi level with positive slope, which is a symmetry-protected topological invariant [13]. In 3D topological insulators, the crossings of energy levels form Dirac cones in the projected energy graphs over the boundary Brillouin zone $T^2$. The edge characterization of the topological indices of a 3D topological insulator is given by the topological invariants ($\nu_1$, $\nu_2$, $\nu_3$) of the three 2D projections of the Brillouin bundle over the three $T^2$ edges and an extra invariant $\nu_0$, which is a genuine 3D topological invariant that discriminates between strong and weak topological insulators. It is possible to show that the boundary characterization of Kane– Mele indices is topologically invariant [14] and equivalent to the bulk characterization [17].

**A landscape of topological matter.**

The basic ingredients that gave rise to the birth of topological insulators — time-reversal invariance and nontrivial topology of energy bands — can be extended to other discrete symmetries [12,18–20] and appear in many different contexts. The search for spectral structures similar to the band spectrum of electrons in a crystal with spin-orbit couplings is now thriving. For instance, topological phases and topological insulators can be obtained by engineering spectral band structures in optical lattices with ultracold atoms. The versatility of these systems can allow the study of strongly correlated topological phases (see the Progress Article by Goldman *et al.* [21]).

There are interesting topological phases even in bosonic systems. That is the case for topological superconductors, in which the role of time reversal is replaced by particle–hole symmetry (see Commentary by Beenakker and Kouwenhoven [22]). In gyroelectric photonic crystals it is also possible to engineer photonic band structures in the microwave spectral regime with nontrivial Chern classes. The dispersion relations of the corresponding chiral edge states have been experimentally observed (see the Commentary by Lu *et al.* [23]).

Although crystalline translation symmetry is instrumental in setting up the topology of the band structure, it is not absolutely necessary. A small amount of disorder does not destroy the topological behaviour of the system, and even a more sophisticated type of symmetry, like those that arise in quasicrystals, might lead to some spectral band structures susceptible of hosting interesting topological effects (see the Commentary by Kraus and Zilberberg [24]).

Once again, the underlying root of all these phenomena is the nontrivial topology of the bands in the energy spectrum. By the same mechanism topological phases of matter can also arise in pure classical systems. For instance, 2D families of coupled gyroscopes and pendula can display nontrivial topological band structures (see the Commentary by Huber [25]).

There are many other phases of matter in which different topological characteristics play a fundamental role, such as Dirac, Weyl and nodal line semimetals, topological crystalline insulators and topological fluids. The subject is in full effervescence.

It is also worth noting that in gauge theories, the topology of the bundle in which the gauge fields are defined is not the only topological factor which is relevant for physical effects; the topological structure of the actual space of gauge fields also plays a role in physical phenomena like gauge anomalies [26,27]. It is therefore possible that it could also play a relevant role in the analysis of new topological phases of matter.

In the dawn of the topological matter revolution, one of the major challenges of pure topology — the famous Poincaré conjecture — has finally been proved [28–30]: any 3D topological space $X$ with $\pi_1(X) = 0$ is topologically equivalent to the $S^3$ sphere. Recalling Gamow's remark about the unforeseeable applications of topology, there is no doubt that, in one way or another, this result will turn out to be helpful in understanding the riddles of nature.

## Acknowledgements

This work was partially supported by the Spanish MINECO/FEDER grant FPA2015-65745-P and DGA-FSE grant 2015-E24/2.

**References**

1. Gamow, G. *Biography of Physics* (Dover, 1961).

2. t'Hooft, G. in *Recent Developments in Gauge Theories* (ed. t'Hooft, G. et al.), (Plenum Press, 1980) pages 117-133.

3. Lachieze-Rey, M. and Luminet, J.-P. *Phys. Rep.* **254,** 135–214 (1995).

4. Luminet, J. P., Weeks, J., Riazuelo, A., Lehoucq, R. and Uzan, J. P. *Nature* **425,** 593–595 (2003).

5. Cornish, N. J., Spergel, D. N., Starkman, G. D. and Komatsu, E. *Phys. Rev. Lett.* **92,** 201302 (2004).

6. Planck Collaboration et al. *Astron. Astrophys.* **571,** A26 (2014).

7. Steenrod, N. E. *The Topology of Fibre Bundles* (Princeton Univ., 1951).

8. Thouless, D. J., Kohmoto, M., Nightingale, M. P. and den Nijs, M. P. M. *Phys. Rev. Lett.* **49,** 405–408 (1982).

9. Hasan, M. Z. and Kane, C. L. *Rev. Mod. Phys.* **82,** 3045–3067 (2010).

10. Qi, X.-L. and Zhang, S.-C. *Rev. Mod. Phys.* **83,** 1057–1110 (2011).

11. Moore, J. E. *Nature* **464,** 194–198 (2010).

12. Kitaev, A. *AIP Conf. Proc.* **1134,** 22–30 (2009).

13. Fu, L., Kane, C. L. and Mele, E. J. *Phys. Rev. Lett.* **98,** 106803 (2007).

14. Berry, M. V. P. Roy. Soc. Lond. A Math. **392,** 45–57 (1984).

15. Mukai, S. *Nagoya Math. J.* **81,** 153–175 (1981).

16. Wang, Z., Qi, X. L. and Zhang, S. C. *New J. Phys.* **12,** 065007 (2010).

17. Kane, C. L. and Mele, E. J. *Phys. Rev. Lett.* **95,** 146802 (2005).

18. Fu, L. and Kane, C. L. *Phys. Rev. B* **76,** 045302 (2007).

19. Schnyder, A. P., Ryu, S., Furusaki, A. and Ludwig, A. W. *Phys. Rev. B* **78,** 195125 (2008).

20. Ryu, S., Schnyder, A. P., Furusaki, A. and Ludwig, A. W. *New J. Phys.* **12,** 065010 (2010).

21. Goldman, N., Budich, J. C. and Zoller, *Nature Phys.* **12,** 639–645 (2016).

22. Beenakker, C. and Kouwenhoven, L. *Nature Phys.* **12,** 618–621 (2016).

23. Lu, L., Joannopoulos, J. D. and Soljačić, M. *Nature Phys.* **12,** 626–629 (2016).

24. Kraus, Y. E. and Zilberberg, O. *Nature Phys.* **12,** 624–626 (2016).

25. Huber, S. D. *Nature Phys.* **12,** 621–623 (2016).

26. Atiyah, M. F. and Bott, R. *Phil. Trans. Roy. Soc. Lond. A* **308,** 523–615 (1982).

27. Asorey, M. and Mitter, P. K. *Ann. Inst. H. Poincare* **45,** 61–78 (1986).

28. Perelman, G. Preprint at https://arxiv.org/abs/math/0211159 (2002).

29. Perelman, G. Preprint at https://arxiv.org/abs/math/0303109 (2003).

30. Perelman, G. Preprint at https://arxiv.org/abs/math/0307245 (2003).